\documentclass[pra,twocolumn,showpacs,preprintnumbers,amsmath,amssymb]{revtex4}

\usepackage{graphicx}
\usepackage{dcolumn}
\usepackage{bm}
\usepackage{mathrsfs}
\usepackage{epsfig}
\newcommand{\eq}[1]{Eq.~(\ref{#1})}
\newcommand{\fig}[1]{Fig.~\ref{#1}}
\newcommand{\be}[1]{\begin{equation}\label{#1}}
\newcommand{\ee}{\end{equation}}
\renewcommand{\vec}{\mathbf}

\begin{document}

\title {Electron stripping and re-attachment at atomic centers using attosecond half-cycle pulses} 
\author{Agapi Emmanouilidou$^{1}$ and T. Uzer$^{2}$}
\address{$^{1}$Institute for Theoretical Science, University of Oregon, Eugene, Oregon 97403-5203\\
$^{2}$School of Physics, Georgia Institute of Technology, Atlanta, Georgia 30332-0430}   
\date{\today}
\begin{abstract}
We investigate the response of two three-body Coulomb systems when driven by attosecond half-cycle pulses: The hydrogen molecular ion and the helium atom. Using very short half-cycle pulses (HCPs) which effectively deliver ``kicks'' to the electrons, we first study how a carefully chosen sequence of HCPs can be used to control to which of one of the two fixed atomic centers the electron gets re-attached. Moving from one electron in two atomic centers to two electrons in one atomic center we then study the double ionization from the ground state of He by a sequence of attosecond time-scale HCPs, with each electron receiving effectively a ``kick'' from each HCP. We investigate how the net electric field of the sequence of HCPs affects the total and differential ionization probabilities.

 \end{abstract}
\maketitle   

\section{Introduction}
Rapid developments in the generation of attosecond time scale laser pulses \cite{Starace} makes investigations of the interaction of multi-electron atoms with ultra-short laser pulses timely. 
The control and manipulation of Rydberg atoms using one or more half-cycle pulses (HCPs), with each HCP of duration $\tau<< T_{n}$ where $T_{n}$ is the classical electron orbital time, has been the subject of intense research lately \cite{tool1}. One of the tools for controlling such states are carefully tailored sequences of HCPs \cite{tool1,tool2}. An interesting aspect of the HCPs is that atoms respond to them very differently than to laser pulses. 
 The response of a single-electron atom to a sequence of HCPs has been thoroughly explored in Ref. \cite{Reinhold2}. 
In addition, for single-electron atoms, the effect that the net field of a sequence of HCPs has on the atomic dynamics \cite{Burg1,Burg2,Burg3} as well as the remarkable control that can be achieved using a chirped train of HCPs \cite{Burg4} are among the problems that have been studied recently.

In the current work, we first address the effect of a sequence of HCPs on the electron in the $H_{2}^{+}$ diatomic ion. The purpose of this section is to refresh concepts of the ``kicked'' one electron dynamics in one atomic center when driven by a sequence of HCPs \cite{Reinhold2,Hahn} and to explore new effects. We begin with the simplest scenario in which  the two atomic centers are so far apart that the electron is initially localized on one
atomic center. The large distance between the fixed atomic centers allows us to make a connection to the response of a single electron atom to HCPs which has been studied intensively. In particular, for HCPs of picosecond duration there is a large number of studies for single ionization of Rydberg atoms (see for example \cite{Burg2,Reinhold2,Hahn,Zhao2,Piraux}).  Very recently, protocols of generating attosecond HCP trains using strong two-color infrared laser pulses have been analyzed \cite{Persson}. 

The other three-body Coulomb system we explore is the He atom where we study the effect of an idealized sequence of HCPs on double ionization both when the net electric field vanishes (even number of kicks) and when it does not (odd number of kicks).

The interaction of multi-electron systems 
with very strong and ultrashort laser pulses is still a wide-open problem. We perform our calculations using the Classical Trajectory Monte Carlo (CTMC) method \cite{Abrimes}. Quite a few studies show
that classical methods can be successful in describing the single ionization of atomic systems when driven by ultrashort and strong laser fields, in very good agreement with quantum mechanical results  \cite{Olson,Duchateau1, Duchateau2}.  However, open issues remain for single ionization of atomic systems such as the effect of the Coulomb interaction on the
ionization process depending on different intensities of the field \cite{Coulomb1,Coulomb2,Coulomb3}.
Another issue concerns the range of parameters over which the classical techniques give the best agreement, for integral as well as differential probabilities, with quantum mechanical studies \cite{Hansen,Illescas,Borbely}. One of the goals of the current classical study is to initiate exploration of
the above issues in the case of the double ionization of the attosecond time-scale driven He.

The parameters we use are chosen to complement recent studies on ionization by strong ultrashort pulses \cite{Darko1,Darko2}. 
 These authors have found that when the electron is driven by two HCPs of alternating sign there is a very high probability for the electron to recombine
to the ground state of H. They have derived analytic expressions, in the framework of the Magnus approximation, for the recombination rates for
 a single-electron atom \cite{Darko1}  and only very recently for two electron atoms \cite{Darko3}. 

The current paper is structured as follows: In Section II we study the effect of one or two HCPs on the electron dynamics in the $H_{2}^{+}$ diatomic ion when the two nuclear centers are
far apart and the electron is initially localized in one of them.  In Section III, we obtain the total and double differential probabilities for the double ionization of the driven He atom. We explore the effect
of a sequence of HCPs when the net electric field vanishes and when it does not and investigate the influence of the electron-electron interaction in the ionization process.

\section{Single-electron diatomic molecular ion driven by strong attosecond pulses} 
In recent years quite a few studies have addressed ionization of single electron atoms
when driven by HCPs, with each HCP with duration $\tau$ much smaller than the time the electron needs to orbit around
the nucleus (Kepler period), $T_{n}$. Let us describe the sequence of HCPs by a field of the form    
$\vec{E}(t)=\sum_{n=1}^{n=N}(-1)^{n}E_{0}f_{n}(t-(n-1) \Delta))\hat{z}$, where the shape of each HCP is given by the function $f_{n}$(t) with $(n-1)\tau<t<n\tau$, and $\Delta$ being the time delay between the HCPs numbered n and n+1.
In these previous studies \cite{Reinhold2,Hahn} it has been well established that the effect of an $N=1$ HCP with $\tau<<T_{n}$ is to deliver a momentum ``kick'', with the energy distribution
of the ionized electron centered around $q^{2}/2-|E_{i}|$ , when $q^2/2-|E_{i}|>0$,with $E_{i}$ being the binding energy of the electron.
If $f_{n}(t)=sin(\pi (t- \Delta)/\tau)$, which is the shape of the HCP we use for all subsequent calculations, then the strength $E_{0}$ of the field is expressed in terms of the momentum ``kick''   
 as $E_{0}=q\pi/(2\tau)$, and the sequence of HCPs we use effectively delivers ``kicks'' of alternating direction. In recent studies \cite{Darko1,Darko2} it was shown that while a single ($N=1$) HCP strips the electron from the hydrogen atom, a subsequent application of a second HCP
 of opposite sign immediately after the end of the first one results in re-attachment of the ionized electron with probability close to one. 
 
Studying the above concepts in a new context we now investigate the effect of a sequence of $N=2$ HCPs
 on the single electron of the $H_{2}^{+}$ ion. We assume that initially the two nuclei are along the z-axis at $R=40$ a.u. and the electron is localized on the left nucleus. Since the
 distance R of the nuclear centers is large it is a very good approximation to consider the initial state of the single electron as that of a hydrogen atom. We take
 as our initial state the 2s state of the hydrogen atom (left nucleus) and consider the two nuclei fixed in space. The initial distribution of the 2s state is given by the        
 microcanonical distribution \cite{Abrimes} $\rho(\gamma)$: 

\begin{equation}
\label{eq:distribution1}
\rho(\gamma) = \mathscr{N}\times
\delta(E_{1}+I_{1})
\end{equation}
with the normalization constant $\mathscr{N}$ and $I_{1}=1/8$ a.u., namely the
ionization energy of the 2s electron. 

\subsection{Single electron ionization in the $H_{2}^{+}$ when driven by an $N=1$ HCP}
\begin{figure}
\scalebox{0.3}{\includegraphics{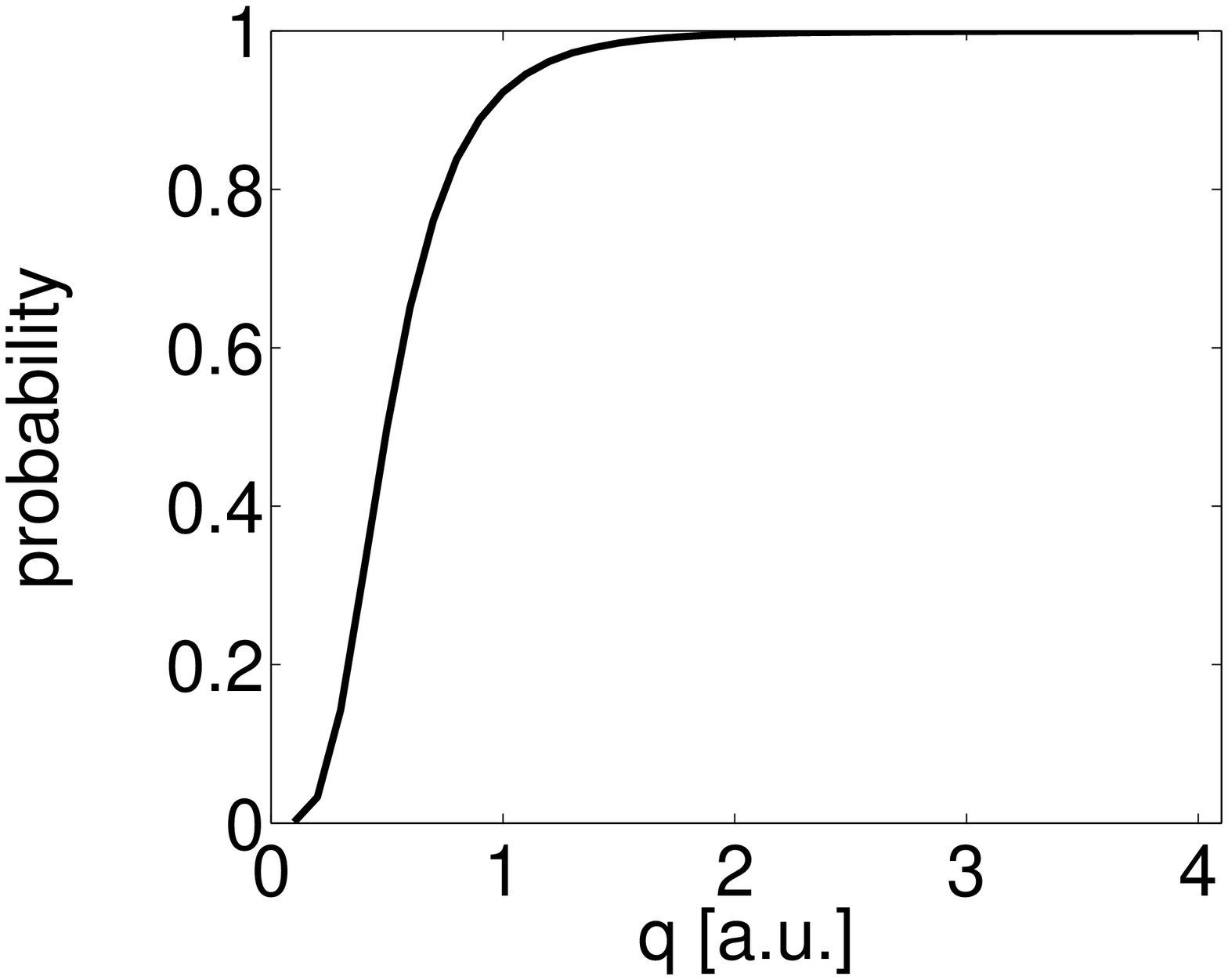}}
\caption{\label{fig1} Probability for ionization of the electron from a 2s state of a hydrogen atom driven by an $N=1$ HCP with $\tau=3$ a.u.}
 \end{figure}

We first study the effect of an $N=1$ HCP on the diatomic ion using the full-three body Hamiltonian with nuclei fixed in space.

\begin{equation}
\label{molecule}
H=p^2/2-1/|\vec{r}-\vec{R_{1}}|-1/|\vec{r}-\vec{R_{2}}|-\vec{r}\cdot \vec{E}(t),
\end{equation}
with $\vec{R}_{1}$, $\vec{R}_{2}$ the position vectors of the two nuclear centers and $\vec{E}(t)$ the field specified above polarized along the direction of the molecular axis z. 
The probability for ionization from the diatomic ion as a function of the strength of the momentum transfer q is shown in \fig{fig1} for a HCP with $\tau=3$ a.u. Since the nuclei are so far apart
it is of course to be expected that our results in \fig{fig1} are exactly the results one would obtain from the single ionization of hydrogen from the 2s state, as is indeed
the case \cite{Darko2}. In our classical calculation with the initial distribution considered in \eq{eq:distribution1} the agreement with quantum mechanical results is better the larger the strength of the field
is and thus the higher it is from the threshold field strength corresponding to over the barrier ionization.  Let us also emphasize that for all the results presented the full Hamiltonian
of \eq{molecule} is used for the propagation in time and our reference to ``kicks'' is only an interpretation of the accurate results obtained with all interactions accounted for.

\subsection{Controlling the atomic center of electron re-attachment in $H_{2}^{+}$ using $N=2$ HCPs}
\begin{figure}
\scalebox{0.3}{\includegraphics{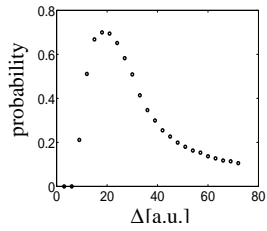}}
\caption{\label{fig3} Probability for attachment in the right nuclear center when the electron is initially localized in the left nuclear center at the 2s state and is then driven by $N=2$ HCPs with $q=2$ a.u. and $\tau=3$ a.u.  }
\end{figure}

Next, we consider the case of $N=2$ HCPs, with $q=2$ a.u. and $\tau=3$ a.u., where we vary the time delay $\Delta$ between the two HCPs. What we are effectively
doing is first ``kicking'' the electron from left to right with a momentum transfer of $q=2$ a.u. and then ``kicking'' it in the opposite direction with a delay $\Delta$. At the end
of the two HCPs we compute the electrons that remain bound, that is, the trajectories for which the energy of the electron given from \eq{molecule} is negative. The electrons are bound either to the
left or right nuclear center which we specify by checking whether $z<0$ or $z>0$ respectively, where z is the coordinate of the electron along the molecular axis, with the 0 of the axis
being at the center of the two nuclei. The results are shown in \fig{fig3}. As expected, when the two HCPs which have opposite direction and equal strength are delivered 
one immediately after the other the electron remains almost completely bound to the left nucleus and the probability to be bound to the right one is zero. However, as the delay between the two pulses increases the probability to find the electron bound to the right nucleus significantly increases. This is reasonable since the longer the delay time,  
the more time the ionized electron from the left nuclear center needs to travel further towards the right nucleus. When the second ``kick'' is received in the opposite direction it reduces
the kinetic energy of the electron ionized from the left nucleus, and depending on the electron's position from the right nuclear center it can result in re-attachment of the electron to the right nucleus. Note that the maximum of the re-attachment probability takes place at a $\Delta$ very close to the time it takes for the electron with a momentum approximately $q=2$ a.u.
to travel from the left to the right nuclear center $R/q=40/2=20$ a.u. Note that our three-dimensional calculation indicates that 
the probability for the electron to be re-attached when one uses HCPs is very large which is to be expected since the ionization probability from the first HCP is very large launching with high probability electrons to the continuum and thus resulting in higher recombination probability to the other nuclear center. 
With rapid experimental advances, this method could be used in the future for determining distances between atomic centers. For related work on how to use attosecond pulses (not HCPs) to observe effects similar to the ones discussed above in diatomic ions see ref. \cite{Bandrauk4,Bandrauk5,Bandrauk6}.

\section{Two electrons driven by strong attosecond pulses} 
We now turn from single electron dynamics in one atomic center to two-electron dynamics in two atomic centers.
In the case of a driven two-electron atom the Hamiltonian is
\begin{equation}
\label{twoH}
H=\frac{p_{1}^2}{2}+\frac{p_{2}^2}{2}-\frac{Z}{r_{1}}-\frac{Z}{r_{2}}+\frac{1}{\left | \vec{r}_{1}-\vec{r}_{2} \right |}+(\vec{r}_{1}+\vec{r}_{2}) \cdot \vec{E}(t).
\end{equation}
In what follows we use as an electric field the same linearly polarized sequence of HCPs as for
the case of the one electron atom in the previous section with the time delay being fixed now and equal to $\tau$ (i.e., effectively we have a sine pulse).

\subsection{Initial phase space distribution for the two-electron atom}

The initial  phase space density
$\rho(\gamma)$ in our classical calculation of the
double ionization of He is given by a product of microcanonical distributions

\begin{equation}
\label{eq:distribution}
\rho(\gamma) = \mathscr{N}
\delta(\epsilon_{1}+I_{1})\delta(\epsilon_{2}+I_{2}),
\end{equation}
with normalization constant $\mathscr{N}$. In the case of He we account for the
electron-electron repulsion in the initial state indirectly through the use of effective charges \cite{Emmanouilidou2}. In the following we present results for two sets of effective charges: a) $I_{1}=I_{2}=Z_{eff}^{2}/2$
where $Z_{eff}=27/16$ and b) $I_{1}=Z_{eff_{1}}^{2}/2$ and $I_{2}=Z_{eff_{2}}^{2}/2$ with $I_{1}$ and $I_{2}$ being
the ionization potentials for the two electrons in the $1s^{2}$ state of He, i.e., $I_{1}=2$ and $I_{2}=0.9$ a.u. The latter choice of effective charges accounts  better for the electron-electron repulsion in the initial state.

\begin{figure}
\scalebox{0.3}{\includegraphics{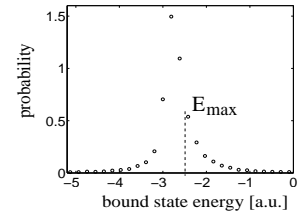}}
\caption{\label{bound}Total energy distribution of the two electrons in the initial state
without a field.}
\end{figure}

\begin{figure}
\scalebox{0.3}{\includegraphics{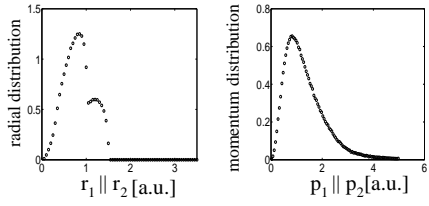}}
\caption{\label{initialproball}Left figure: Radial distribution as a function of the radial electronic coordinate; right figure: momentum distribution as a function of the electronic momentum. Since we use different
effective charges the two electrons are not equivalent and so we plot as a function of the radial coordinate
of electron 1 or 2 ($r_{1}||r_{2}$) and similarly for the momentum distribution. }
\end{figure}

\begin{figure}
\scalebox{0.3}{\includegraphics{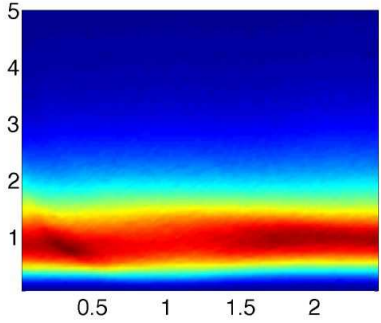}}
\caption{\label{evolution} The field-free evolution in time of the initial momentum distribution. At time zero it is identical to the momentum distribution shown in \fig{initialproball}.}
\end{figure}

A few more remarks on the construction of the initial state are in order: Our initial state is not an eigenstate of the driven two-electron Hamiltonian in \eq{twoH}, and it can thus
``auto-ionize"  even without the presence of the external field. We avoid the latter problem in the following
way: i)  We first generate the initial conditions for each electron independently using the
microcanonical distribution for a one electron atom with charge $Z_{eff_{1}}$ or $Z_{eff_{2}}$ for electron 1 and 2, respectively. We then use the initial conditions we have just generated to obtain the total energy of the two electron atom from \eq{twoH} by setting the field equal to zero. The initial total energy distribution
of the two electrons is shown in \fig{bound}, where we see that there is a long tail on both
sides of the ideal value of -2.9 a.u. We cut the tail off by introducing two parameters $E_{min}$, $E_{max}$ such that \cite{bound}: 

\begin{equation}
\label{cutoff}
\frac{\int_{E_{min}}^{E_{max}}E\rho(E) dE }{\int_{E_{min}}^{E_{max}}\rho(E)}=E_{p},
\end{equation}
with $E_{p}$ the most probable energy and $\rho(E)$ the energy distribution. In our calculations $E_{max}=-2.51$. ii) In addition, we evolve the two-electron atom freely (i.e., without an external field and using 
the full three-body Coulomb Hamiltonian) and discard the trajectories for which the total two electron energy becomes positive during the field-free propagation. We freely propagate the system for times
twice the Kepler period, with the latter being approximately 2.2 a.u. for the He ground state. In this latter step we find that the fraction of trajectories labeled as ``auto-ionizing'' and thus discarded in our simulation is small.  After these steps, we obtain an initial state distribution which has the radial and momentum distribution shown in \fig{initialproball}. Finally, we also find
that the field-free evolution of the initial momentum distribution changes little for times comparable to the Kepler period (see \fig{evolution}) and we can thus consider our ensemble of initial conditions as approximately stable for all practical considerations.

\subsection{Computation of the doubly ionizing trajectories}
 We determine doubly ionizing trajectories due to the HCP's as follows: when the pulse is switched off  we check the total energy of the three-body Coulomb system. If the energy
 is negative then the driving field has not transfered enough energy to cause double ionization and we thus label those trajectories as non-doubly ionizing trajectories. If the energy is greater than zero then it can be the case that these trajectories will lead to single or double ionization. To decide if these latter trajectories are singly or doubly ionized we continue to monitor them in time by propagating the freely evolving three-body Coulomb system until the asymptotic regime is reached.  If at that time $\epsilon_{i}=p_{i}^2/2-Z/r_{i}$, with $Z=2$, for each electron are both positive we label these trajectories as double ionizing otherwise as non double ionizing. Note, that the above described method does not allow us 
 to separate the singly ionized from the bound trajectories. The reason is that with the process described above the $\epsilon_{i}=p_{i}^2/2-Z/r_{i}$ reach constant values only asymptotically but when the system is propagated for so long the classical nature of our calculation can cause bound trajectories to become ``auto-ionized'' and thus artificially contribute to the single
 ionization probability.   
 To compute the double ionization probability we first find the number of double ionizing trajectories and normalize with respect to the total number
of trajectories that have been propagated. The single and double differential probabilities presented in this paper
use the asymptotic values of the quantities plotted. We believe that even though we account for the electron-electron correlation in the initial state through effective charges this simple initial state
captures accurately the essential final correlations as has been shown in other cases \cite{Emmanouilidou2}.  

\subsection{Single and double differential probabilities for double ionization}
\begin{figure}
\scalebox{0.35}{\includegraphics{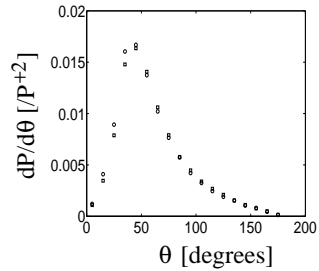}}
\caption{\label{anglescomp} Distribution of the inter-electronic angle for a laser pulse with $q=3$ a.u. 
and $\tau=0.1$ a.u. . Results for $N=1$ HCP are shown: The $\blacksquare$ refer to $I_{1}=I_{2}$ while the $\circ$ refer to $I_{1}\neq I_{2}$. }
\end{figure}

\begin{figure}
\scalebox{0.35}{\includegraphics{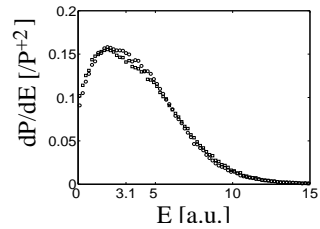}}
\caption{\label{energiescomp}Energy distribution normalized over the double ionization probability for a laser pulse with $q=3$ a.u.
and $\tau=0.1$ a.u. . Results for $N=1$ HCP are shown: The $\blacksquare$ refers to $I_{1}=I_{2}$ while the $\circ$ refers to $I_{1}\neq I_{2}$. }\end{figure}

  Following the procedure previously described, we construct the initial state distribution both for equal
  and different effective charges and find the trajectories that doubly ionize after $N=1$ HCP. We find
  that both initial state distributions yield similar results both for the energy as well as the 
inter-electronic angular distribution, as can be seen in \fig{anglescomp} and  \fig{energiescomp}. 
 In the following we present results for the initial state distribution that corresponds to $I_{1}\neq I_{2}$, 
In addition, for the single differential probabilities presented in this paper for $I_{1}\neq I_{2}$
we take the average of the differential probabilities of 
the two electrons unless otherwise specified.

\begin{figure}
\scalebox{0.35}{\includegraphics{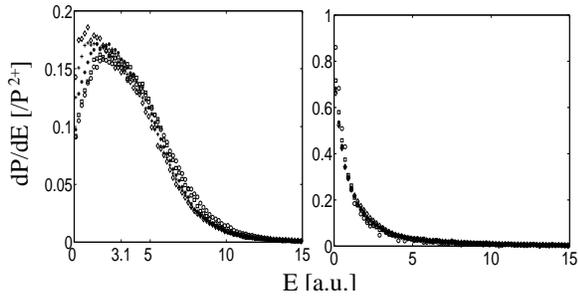}}
\caption{\label{fig8}Energy distribution for a laser pulse with $q=3$ a.u.
and $\tau=0.1$ a.u. Left panel: results for an odd number of HCPs with $N=1$ $(\bullet)$, $N=3$ $(\blacksquare)$, $N=5$ (*), $N=7$ (+), $N=9$  $(\diamond)$; Right panel:  
results for an even number of HCPs  with $N=2$ $(\bullet)$, $N=4$ $(\blacksquare)$, $N=6$ (*), $N=8$ (+), $N=10$ $(\diamond)$.}
\end{figure}

In \fig{fig8}, we present results for the energy distribution
 for odd or even number of HCPs for $q=3$ a.u. and $\tau=0.1$ a.u., with $\tau$ much smaller than the Kepler period $T_{n}$ of an electron in a hydrogenic atom, $T_{n}= 2\pi n^{3}/Z_{eff}^2$. Let us note that while an even number of HCPs corresponds to a zero
  time integral of the complete pulse and it can thus be produced, the odd number of HCPs corresponds to a non-zero time integral of the pulse and can
  thus not be produced \cite{Madsen}. However, what can be produced instead of an $N=1$ HCP is a very short and strong first half cycle followed by a much weaker and much longer second half to compensate \cite{tool2}.   
 While the interaction of 
 each of the electrons with the laser pulse is very strong, the electron-electron interaction as well as the interaction of each of the electrons
 with the nucleus is not negligible. If the latter were of no importance, double ionization would not take place, since an even number of HCPs transfers zero net momentum. This has already been noted for the ionization of the driven hydrogen, see Ref. \cite{Darko2}.  There it was pointed out that one can think of the process as a sequence of $\delta$ kicks, at times equal with an integer multiple of $\tau/2$,
 with opposite signs (HCPs of opposite sign) where the interaction of the electron with the nucleus in between the kicks can not be neglected.  
   For an even number of HCPs it is no surprise that on average the momentum transfer is small and the kinetic energy acquired is less than the binding energy of the two-electron atom, $E_{i}$, with the energy distribution peaking at energies close to zero.
   However, when an odd number of HCPs are applied then the net momentum transfer is $q$ and since the kinetic energy for each electron is greater than the electron's binding energy, $q^{2}/2>|E_{i}|/2$, 
one expects the energy distribution to peak around $q^{2}/2-|E_{i}|/2\approx 3.1$ a.u. if the electron-electron interaction plays no role. From \fig{fig8} we see that the electron distribution peaks at an energy smaller than 3.1 a.u. 
  The latter effect can not be due to our approximate initial state since this latter approximation can only
  cause the energy distribution to peak at a higher energy (effectively, smaller $E_{i}$).   As we increase the number of odd HCPs the shift of the energy distribution to smaller energies becomes more substantial, although one can clearly see a ``shoulder'' structure around $E=3.1$ a.u. 
  This shift to smaller energies for an increased number of odd HCPs must be due to the increasing significance of the electron-electron repulsion in the dynamics
  of the doubly ionizing trajectories for the application of longer pulses. Indeed, for an even number of HCPs where the electron-electron interaction
  is more important (see also next paragraph) the energy distribution is peaked around small energies.

\begin{figure}
\scalebox{0.35}{\includegraphics{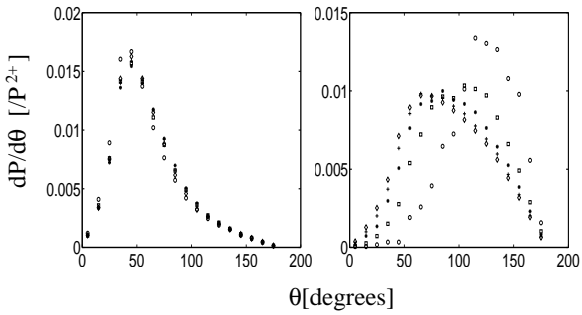}}
\caption{\label{fig9} Distribution of the inter-electronic angle for a laser pulse with $q=3$ a.u.
and $\tau=0.1$ a.u.  Left panel: results for an odd number of HCPs with $N=1$ $(\bullet)$, $N=3$ $(\blacksquare)$, $N=5$ (*), $N=7$ (+), $N=9$  $(\diamond)$; Right panel:  
results for an even number of HCPs  with $N=2$ $(\bullet)$, $N=4$ $(\blacksquare)$, $N=6$ (*), $N=8$ (+), $N=10$ $(\diamond)$.}
\end{figure}

In \fig{fig9}, we show the distribution of the inter-electronic angles for odd as well as even numbers of HCPs. We see that for
an odd number the distribution peaks at approximately $45^{\circ}$ while for an even number of HCPs its peak shifts from larger to smaller angles
as the number of HCPs increases. The fact that the inter-electronic angle between electrons
that have received an even number of kicks is larger than the angle between those which have been kicked an odd number of times is to be expected:
In the former case the electrons are much slower (compare the right and left panels in \fig{fig8}) and thus the electron-electron repulsion
is much more pronounced as compared to the electrons that escape after an odd number of cycles. This will be further illustrated in the following when we present
single and double differential probabilities in angle.

  
 In Table I we present the probability for double ionization
as a function of the number of HCPs.  We find that the double ionization probability follows the same pattern as the single ionization of H
\cite{Darko2}. Namely, as the number of even HCPs increases the probability for double ionization increases whereas the probability for double ionization
decreases with increasing number of odd HCPs (Table I shows that the probability for both electrons to remain bound after
a large number of even cycles is significant.) So as for the case of a single electron atom 
for the case of two electron atoms the electrons are ``stripped'' from the atom when driven by an odd number of HCPs while an even number of HCPs causes the two electrons to re-attach to the atomic center \cite{Brigsnew}. Also, in \fig{fig22} we show how the double ionization probability changes as a function of the momentum transfer q for the case of $N=8$ and $N=9$ HCPS.

\vspace{1cm}
\begin{tabular}{|c|c|c|c|} \hline
\# half-cycles &Prob& \# half-cycles & Prob \\ \hline
    1                          &   0.70                 &                                    & \\ \hline
                                 &                    &   2  & 0.0029 \\  \hline
     3                                 & 0.67                  &        &\\ \hline
                                 &                    &4& 0.0177\\ \hline
      5                        &  0.58                   &  &\\ \hline
                                 &                    &6&0.053\\ \hline
       7                       &   0.51                &   &\\  \hline
                                 &                   &8&0.1\\ \hline
        9                       &    0.48              & &\\ \hline
                                  &                  &10&0.15\\ \hline     
                                  \end{tabular}                                                                                       
\vspace{1cm}
  
  In addition, we show in \fig{fig12} and \fig{fig13} the probability density of the momentum along the z-axis (see Ref.\cite{Emmanoui}
  for the definition of the classical probability density). It is clear that the effect of the field on each of the electrons is a kick received during each of the HCPs. We emphasize that all our results are obtained by propagating with the full three-body Hamiltonian under the influence of an odd or even number of HCPs of total duration $N\tau$, see \eq{twoH}. Referring to ``kicks'' is only an interpretation consistent with
  the results obtained accounting for all Coulomb interactions and accounting for the effect of the finite total duration of the external field applied. The probability density of the momentum along the z axis clearly illustrates that the HCPs applied are very strong.   
    
     
In \fig{fig15} and \fig{fig16} we show the double angle differential probability densities. The angles in question refer to the angle that each of the electrons makes with the polarization axis z.  
We find that when the two electrons are driven by an odd number of HCPs they escape to the continuum almost antiparallel to the electric field (which is the
direction of the force the field exerts on each of the electrons) with a small inter-electronic angle. When the number of cycles is even the two electrons
escape with a much larger inter-electronic angle with one electron at a small angle with respect to the polarization axis.

Finally, in \fig{fig21} and \fig{fig22} we plot the double differential probability with respect to the magnitude of the momenta of electrons 1 and 2.  
Both for the case
of odd and even number of cycles we find that as we increase the number of HCPs more electrons escape with differing momenta, that is, the probability for 
one fast and one slow electron increases with the number of HCPs. Also a comparison between the $N=9$ and $N=10$ HCPs case clearly shows 
that the escape of the two electrons with different momenta is much more pronounced for the case of $N=10$ HCPs. In \fig{fig23} we plot the average
momentum, radius, and relative direction of the momentum and position, $\vec{r} \cdot \vec{p}/(rp)$, for each electron for the case of $N=9$ and $N=10$ HCPs for all doubly ionizing trajectories. We find that at the end of the $N=10$ HCPs the electrons are closer to the nucleus (smaller radii) compared to the $N=9$ case.
Another difference is that while for the case of $N=9$ HCPs the two electrons follow the external field during the time that it is switched on, for the case of $N=10$ HCPs the electrons strongly interact with the nucleus (very small value of the radii) before they follow the driving field.

   In this section we have presented classical results regarding the double ionization of the ground state of the He atom with HCPs with the duration $\tau$ of each 
   HCP being much shorter than the Kepler period. In addition, all our calculations for single and double differential probabilities have been performed
   for values of the peak of the external field $E_{0}$ that are much higher than the threshold value for over the barrier ionization. So our choice of peak strength
    and duration of the pulse is such that the classical calculations would be most reliable. We expect
    that our classical calculations accurately capture the essential features of the double ionizing trajectories particularly of those where the electrons escape with larger energies.
    Our classical results are presented for the $n=1$ state of the He atom using parameters of the external field that correspond to a peak intensity of 
   $7.8 \times 10^{20}$ W/cm$^2$ and an effective frequency, $\pi/\tau$, of 31 a.u. which are well beyond current experimental capabilities. However, the results presented (with an appropriate scaling of the transfer q and the time $\tau$)
   should also hold for two electrons much less bound in the initial state as is the case for the doubly excited resonant states of He which can have lifetimes of picoseconds and are thus within current experimental capabilities.    
 
In summary, we have used  a classical calculation to show how a sequence of HCPS can be used to ``strip'' and re-attach one electron to two atomic centers or two electrons to one atomic center. We have presented results for the single and double differential probabilities for the double
ionization of helium from the ground state when driven by HCPs of attosecond
time scales. We have shown that for an even number of attosecond HCPs the effect of the nucleus on double ionization is pronounced and can not be neglected. We anticipate that as quantum calculations become available a direct comparison with our classical results will better illustrate the regime of validity of the
classical calculations for single and double differential probabilities.

\clearpage

\begin{figure}
\scalebox{0.45}{\includegraphics{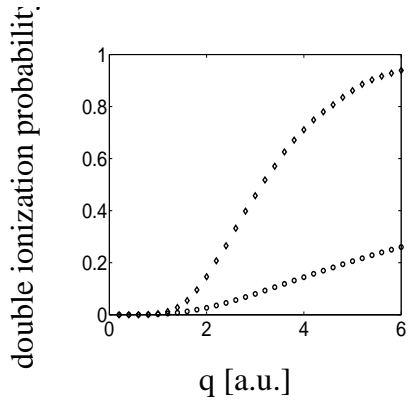}}
\caption{\label{fig25}Double ionization probability as a function of the momentum transfer $q$ for $N=8$ $\circ$ and $N=9$ $\diamond$ HCPs.}
\end{figure}

\begin{figure}
\scalebox{0.35}{\includegraphics{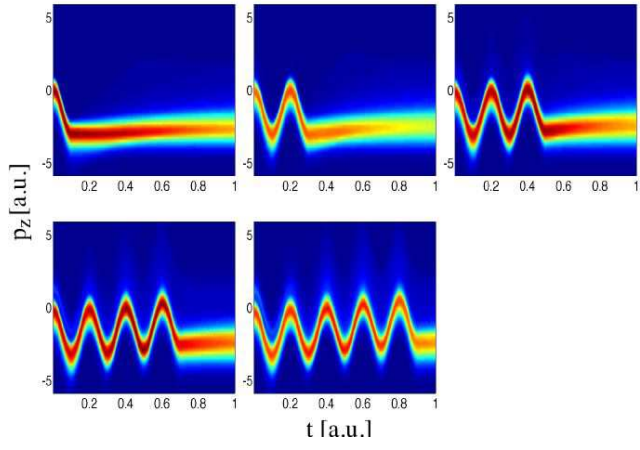}}
\caption{\label{fig12}Probability density of momentum along the z-axis for $q=3$ a.u. and $\tau=0.1$ a.u. Results for an odd number of HCPs with $N=1,3,5,7$ and $N=9$ from left to right.}
\end{figure}

\begin{figure}
\scalebox{0.35}{\includegraphics{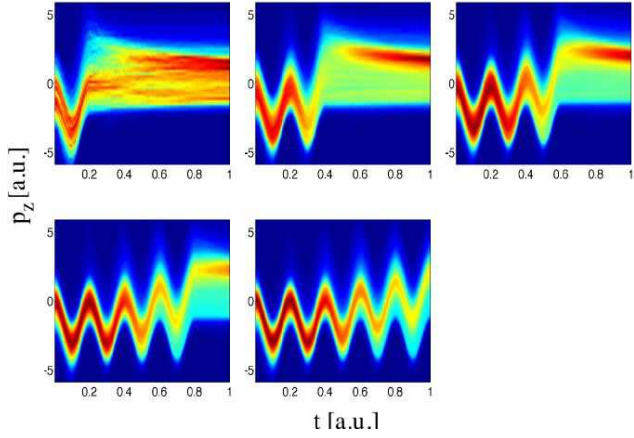}}
\caption{\label{fig13}Probability density of momentum along the z-axis for $q=3$ a.u.  and $\tau=0.1$ a.u. Results for an even number of HCPs with $N=2,4,6,8$ and $N=10$ from left to right.}
\end{figure}

\begin{figure}
\scalebox{0.35}{\includegraphics{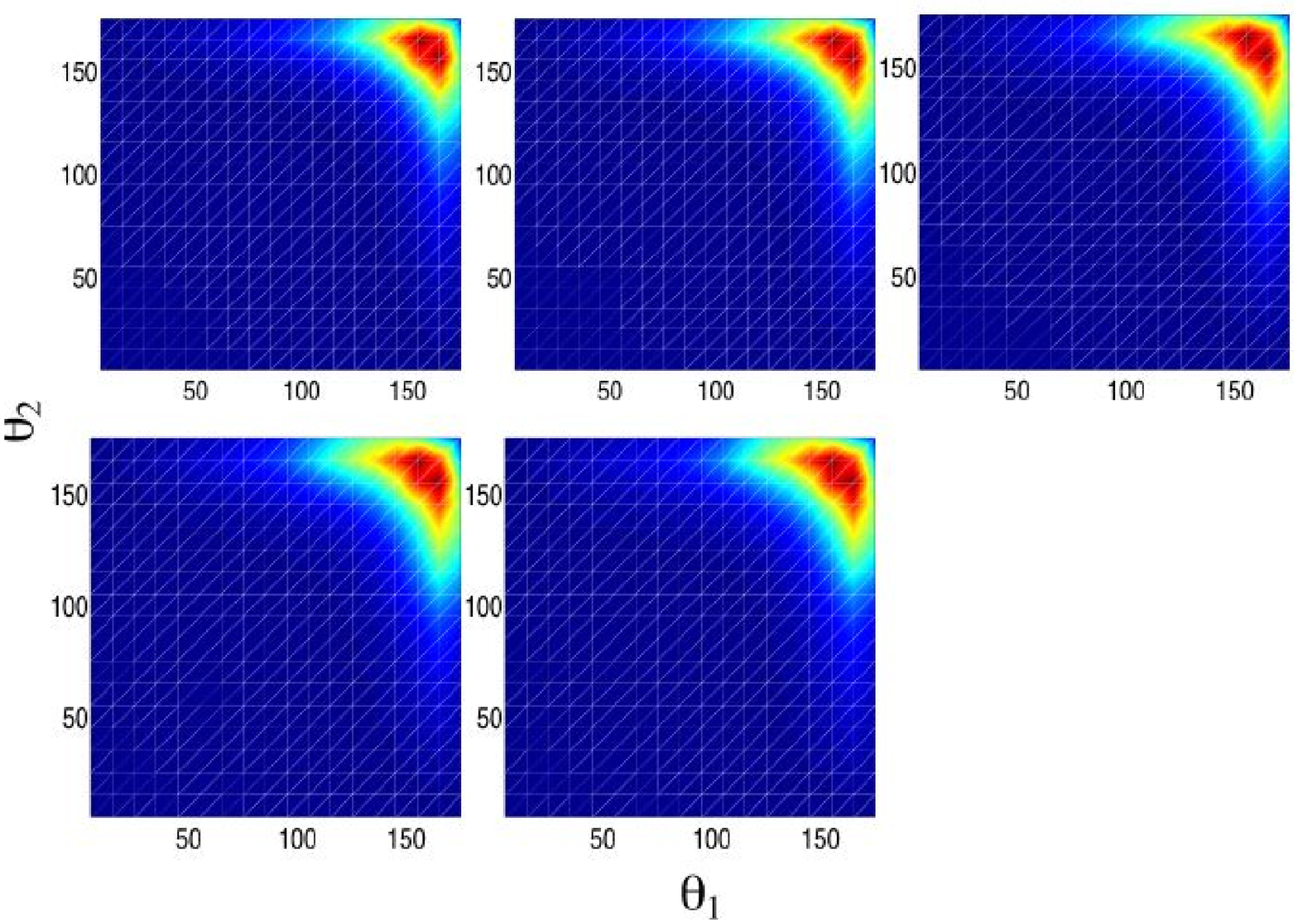}}
\caption{\label{fig15}Double angle differential probability density for a laser pulse with $q=3$ a.u.
and $\tau=0.1$ a.u. The angles $\theta_{1}$ and $\theta_{2}$ are the angles of the two electrons with respect to the direction of the field. $\theta_{i}=0$ corresponds to a direction parallel to the field. Results 
for an odd number of HCPs with $N=1,3,5,7$ and $N=9$ from left to right. Note that the results have been symmetrized with respect to the diagonal $\theta_{1}=\theta_{2}$.}
\end{figure}

\begin{figure}
\scalebox{0.35}{\includegraphics{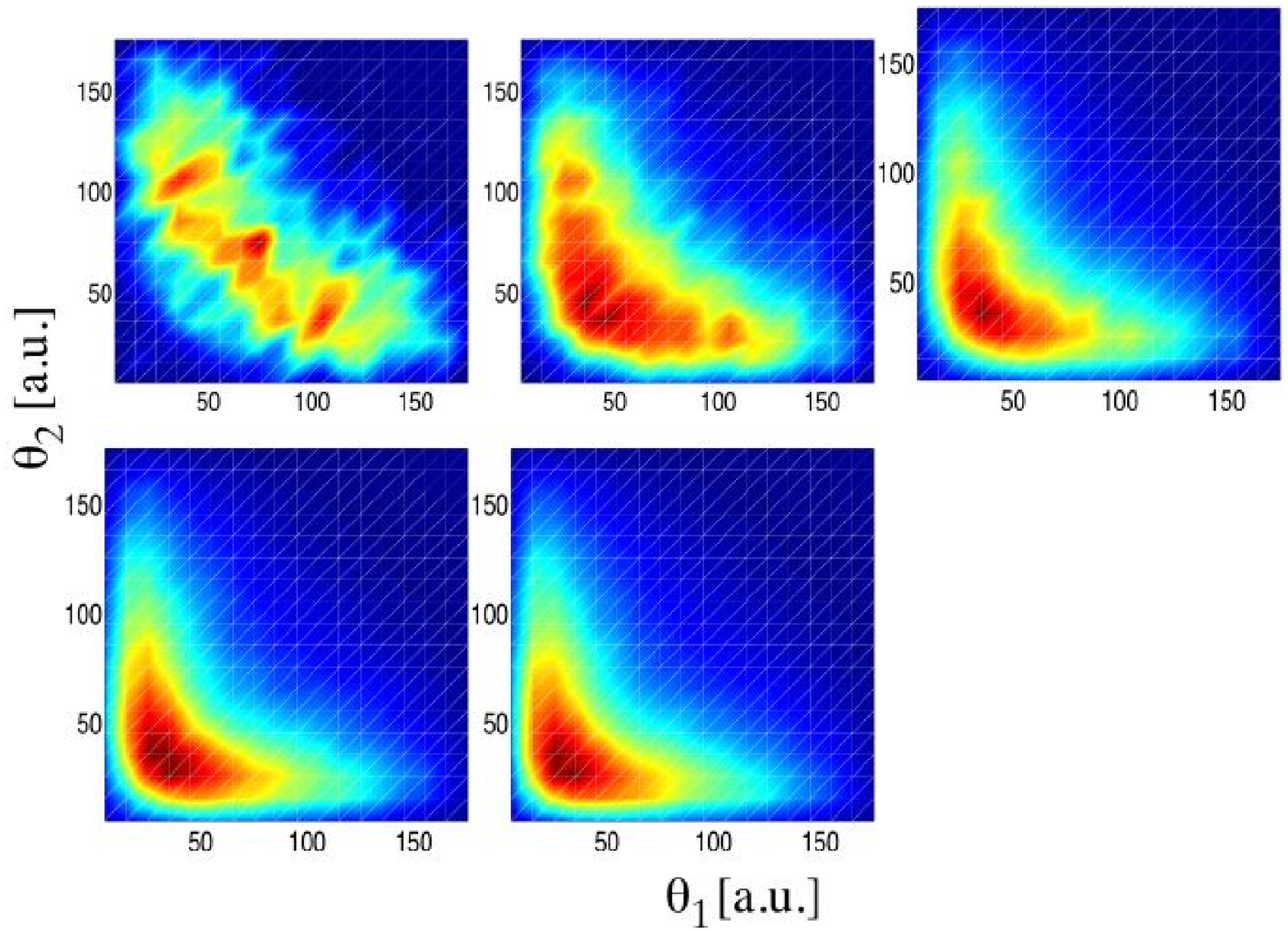}}
\caption{\label{fig16}Double angle differential probability density for a laser pulse with $q=3$ a.u.
and $\tau=0.1$ a.u. The angles $\theta_{1}$ and $\theta_{2}$ are the angles of the two electrons with respect to the direction of the field. $\theta_{i}=0$ corresponds to a direction parallel to the field. Results 
or an even number of HCPs with $N=2,4,6,8$ and $N=10$ from left to right. Note that the results have been symmetrized with respect to the diagonal $\theta_{1}=\theta_{2}$. (The graph for $N=2$ is not smooth due to the lower statistics).}
\end{figure}

\begin{figure}
\scalebox{0.45}{\includegraphics{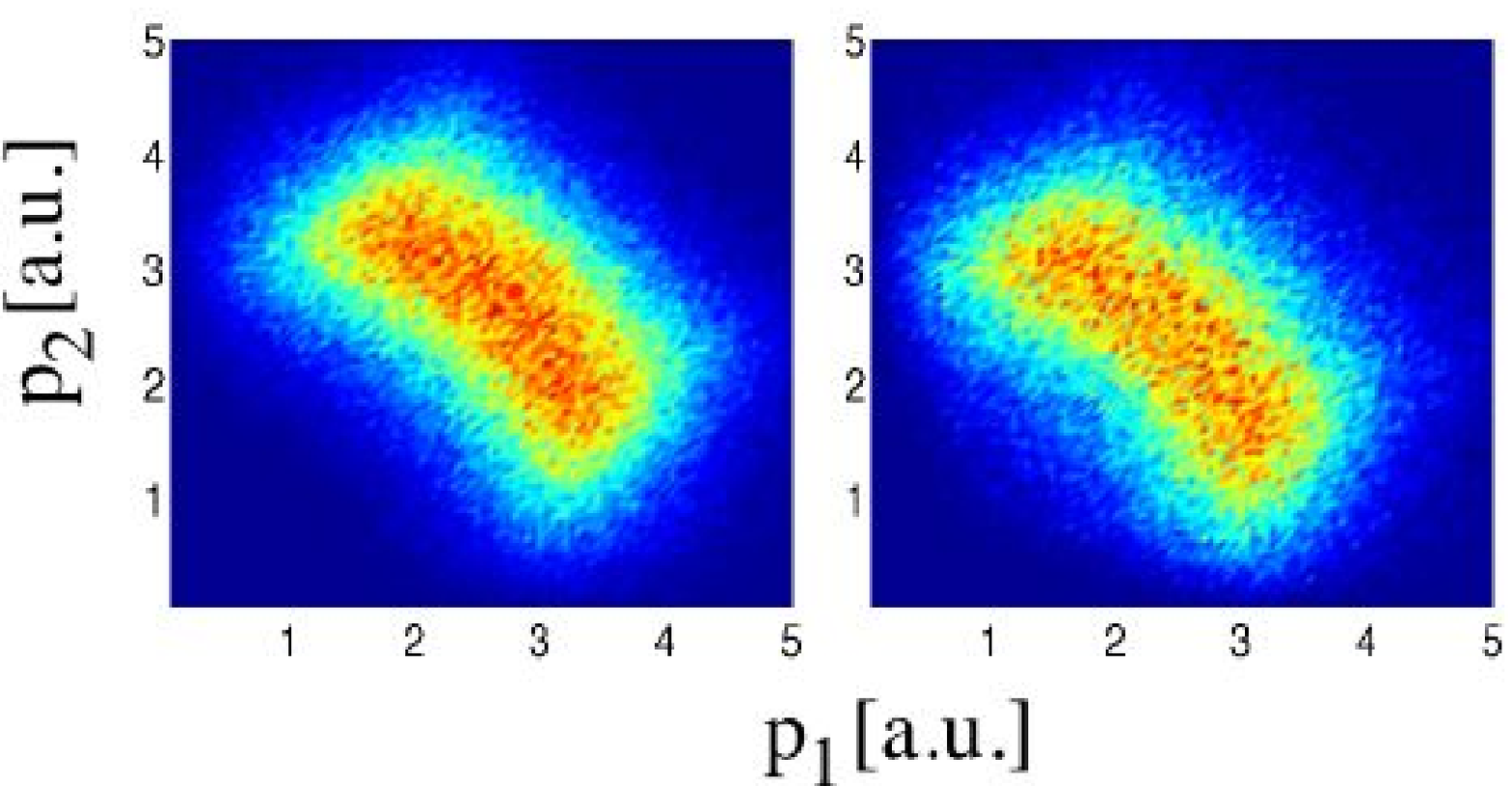}}
\caption{\label{fig21}Double momentum differential probability density for the momentum $p_{1,2}$. The left figure corresponds to $N=1$ HCP and the right one corresponds to $N=9$ HCPs. Note that the results have been symmetrized with respect to the diagonal $p_{1}=p_{2}$.}
\end{figure}

\begin{figure}
\scalebox{0.45}{\includegraphics{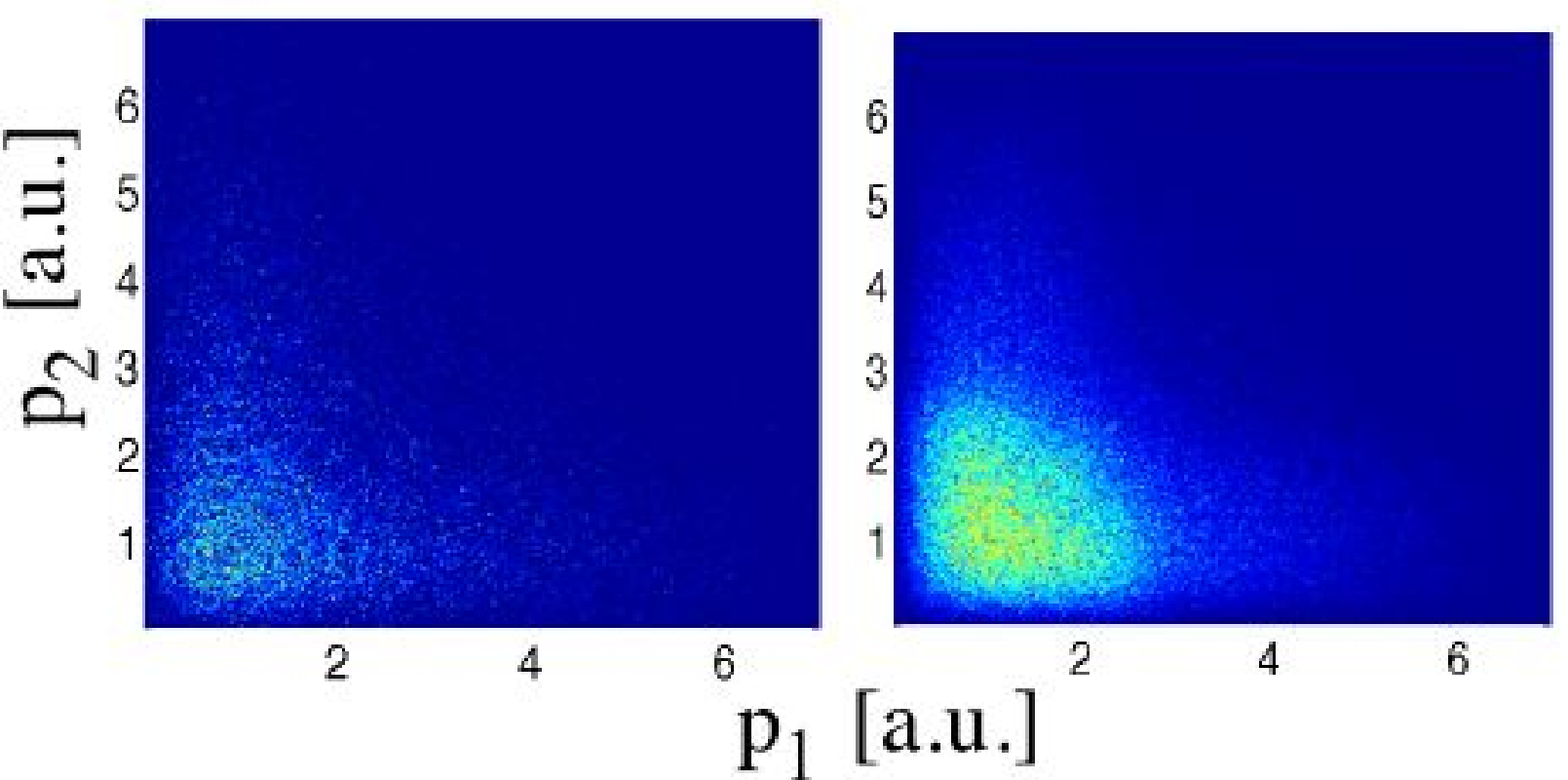}}
\caption{\label{fig22}Double momentum differential probability density for the momentum $p_{1,2}$. The left figure corresponds to $N=2$ HCPs, the right one corresponds to
$N=10$ HCPs. Note that the results have been symmetrized with respect to the diagonal $p_{1}=p_{2}$.}
\end{figure}

\begin{figure}
\scalebox{0.5}{\includegraphics{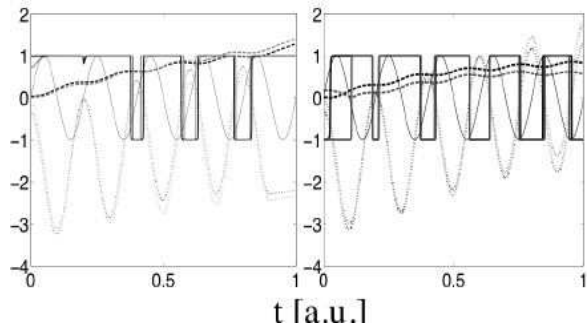}}
\caption{\label{fig23}Left panel for $N=9$ HCPs and right panel for $N=10$ HCPs: Average over all double ionizing trajectories of the radius of electrons 1 (black dashed line) and 2 (gray dashed line), of the relative direction between
the position and the momentum of electron 1 (black solid line) and electron 2 (grey solid line), the momenta parallel to the field for electron 1 (black dotted line) and electron 2 (grey dotted line). The thin solid line represents $sin(\pi/\tau t)$. }
\end{figure}

\end{document}